\documentclass[conference]{IEEEtran}

\usepackage{blindtext, graphicx}

\usepackage{todonotes}
\usepackage[figuresright]{rotating}
\usepackage{amsmath,amssymb}
\usepackage{graphicx}
\graphicspath{ {images/} }
\usepackage{multirow}
\usepackage{url}
\usepackage{verbatim}
\usepackage{subfigure}
\usepackage{multirow}
\usepackage{soul,color,colortbl}
\usepackage{booktabs}
\usepackage{upgreek}
\usepackage{tabulary}
\usepackage{bm}
\usepackage[section]{placeins}
\usepackage{tabularx}
\usepackage[noend]{algpseudocode}
\usepackage[ruled,vlined,linesnumbered]{algorithm2e}
\usepackage{tikz}
\usepackage{adjustbox}                      
\usepackage{enumitem}
\usepackage{array}                              
\usepackage{tabularx}  
\usetikzlibrary{positioning,arrows.meta,decorations.text,calc}
\usetikzlibrary{shapes.multipart}
\usepackage{booktabs} 
\usepackage[english]{babel}
\usepackage[utf8]{inputenc}
\usepackage{amssymb}
\usepackage{enumitem}
\usepackage{verbatim}
\usepackage{pgfplots}
\pgfplotsset{width=7cm,compat=1.8}
\usetikzlibrary{patterns}
\usepackage{verbatim}
\usepackage{pgfplots}
\pgfplotsset{width=7cm,compat=1.8}
\usetikzlibrary{patterns}

\begin{document}

\title{Finding Rats in Cats: Detecting Stealthy Attacks using Group Anomaly Detection}

\author{
\IEEEauthorblockN{Aditya Kuppa\IEEEauthorrefmark{1}\IEEEauthorrefmark{2},
Slawomir Grzonkowski\IEEEauthorrefmark{1}, 
Muhammad Rizwan Asghar\IEEEauthorrefmark{3}, and
Nhien-An Le-Khac\IEEEauthorrefmark{2}}
\IEEEauthorblockA{
\IEEEauthorrefmark{1}Symantec Corporation, Ireland\\
\IEEEauthorrefmark{2}School of Computer Science, University College Dublin, Ireland\\
\IEEEauthorrefmark{3}School of Computer Science, The University of Auckland, New Zealand\\
aditya.kuppa@ucdconnect.ie, slawomir\_grzonkowski@symantec.com, r.asghar@auckland.ac.nz, an.lekhac@ucd.ie
}
}

\maketitle

\begin{abstract}

Advanced attack campaigns span across multiple stages and stay stealthy for long time periods. 
There is a growing trend of attackers using off-the-shelf tools and pre-installed system applications (such as \emph{powershell} and \emph{wmic}) to evade the detection because the same tools are also used by system administrators and security analysts for legitimate purposes for their routine tasks. 
Such a dual nature of using these tools makes the analyst's task harder when it comes to spotting the difference between attack and benign activities.
To start investigations, event logs can be collected from operational systems; however, these logs are generic enough and it often becomes impossible to attribute a potential attack to a specific attack group.

Recent approaches in the literature have used anomaly detection techniques, which aim at distinguishing between malicious and normal behavior of computers or network systems. 
Unfortunately, anomaly detection systems based on point anomalies are too rigid in a sense that they could miss malicious activity and classify the attack, not an outlier.
Therefore, there is a research challenge to make better detection of malicious activities. 
To address this challenge, in this paper, we leverage Group Anomaly Detection (GAD), which detects anomalous collections of individual data points. 

Our approach is to build a neural network model utilizing Adversarial Autoencoder (AAE-$\alpha$) in order to detect the activity of an attacker who leverages off-the-shelf tools and system applications.
In addition, we also build \textit{Behavior2Vec} and \textit{Command2Vec} sentence embedding deep learning models specific for feature extraction tasks. 
We conduct extensive experiments to evaluate our models on real world datasets collected for a period of two months. 
Our method discovered 2 new attack tools used by targeted attack groups and multiple instances of the malicious activity. 
The empirical results demonstrate that our approach is effective and robust in discovering targeted attacks, pen-tests, and attack campaigns leveraging custom tools.

\end{abstract}


\IEEEpeerreviewmaketitle

\section{Introduction}

Today, with the increase in deployments of security controls inside enterprises, attackers are relying on tools already installed on the system for their attack campaigns \cite{istr}. 

Different system tools are leveraged by attackers and cybercriminals alike to achieve their goal \cite{istr}. 
For example, tools like \textit{net} and \textit{powershell} can be used to discover information about the target environment, establish remote connections, run scripts to move laterally inside the victim environment, and exfiltrate sensitive data. 
Recent attacks against the Democratic National Committee (DNC) used \textit{powershell} for lateral movement and discovery. 
The Odinaff group, which attacked SWIFT systems used in the financial services and banking sectors, used \textit{mimikatz} tool to dump user passwords from memory. 
The network scanner allowed the group to identify other computers in the same local network. 
The dumped credentials were then used with \textit{powershell} to start a new process on one of the identified remote computers.
\begin{table*}[!htb]
    \centering

    \caption{Each row represents command line executions in a Windows Session, Colour of row indicates whether  the session is an \textbf{\textcolor{red}{attacker}} session or a   \textbf{\textcolor{blue}{normal}} admin/user session.}
    \label{tab:command_session}
    
\begin{tabularx}{\linewidth}{ |r| X| }
\hline 
    \textbf{Session ID} & \textbf{Commands Executed} \\ \hline
  \textcolor{red}{ S-1-2-331-21} &
    
\textcolor{red}{
\scriptsize 
     \tt{cmd hostname,whoami query user ipconfig -all ping www.google.com }
     \tt{net user, net view /domain tasklist /svc }
     \tt{ netstat -ano | find \%TCP\%  msdtc [IP] [port]}
     }\\
     \hline 
     \textcolor{blue}{ S-1-2-331-22}
     & 
     \textcolor{blue}{
     \scriptsize 
     \tt{ruby.exe" "tools\%ruby\%ridk\%current\%bin\%rspec/cookbooks/configservice}
     \tt{1.exe" "--url" "ssh://access/ssh?team=}
     \tt{iexplore.exe" SCODEF:XXXX CREDAT:XXXXXX /prefetch:X}
     \tt{ping [IP] monitor.log }
     }\\
     \hline 
     \textcolor{blue}{ S-1-2-331-23}
     & 
     \textcolor{blue}{
     \scriptsize 
     \tt{chrome.exe", cmd.exe /K RSSFeeds.bat}
     \tt{cmd.exe" /C "backupupgrade.bat}
     \tt{powershell.exe"Bypass-EncodedCommand JABwACAAPQAgAFMAdA..}
     \tt{powershell.exe iex (Text.Encoding..}
     }
     \\
  \hline 
     \textcolor{blue}{ S-1-2-331-24}
     & 
     \textcolor{blue}{
     \scriptsize 
     \tt{inventory.exe "01-run.xml"}
     \tt{winword.exe" /n "downloads\%dada.doc" /o ""}
     \tt{installer.exe" -o install -ip [localip] -u}
     \tt{appdatalocaltemptmp0158.exe", /c net config workstation}
     }
       \\
  \hline 
   \textcolor{red}{ S-1-2-331-25}
     & 
     \textcolor{red}{
\scriptsize 
     \tt{setuptestcenterwin.lax" "appdata\%local\%temp\%ladaesda.tmp,
     \%Temp\%EWH.bat, cmd.exe /Q /c powershell -nop -w hidden -encodedcommand JABzAD0AT,
     excel.exe" /e, appdata\%roaming\%tmpxxx.exe,
    appdatalocaltemp\%tmpxxxx.exe", /c net config workstation}}
     \\
     \hline 
      \textcolor{blue}{ S-1-2-331-26}
     & 
     \textcolor{blue}{
\scriptsize 
     \tt{cmd.exe/c""apps\%eclipse\%lunaee\_4\_4\_1\%eclipse\%lunaee.bat,
     ipsec.exe, ping -n 1 [IP], "cmd",
     query.exe" user, "cmd.exe" /c whoami, whoami.exe" /user
     temp\%update.exe} }
     \\
     \hline 
     
\end{tabularx}

\end{table*}

\begin{figure*}[htp]
  \centering

  \subfigure[]{\includegraphics [width=0.25\textwidth]{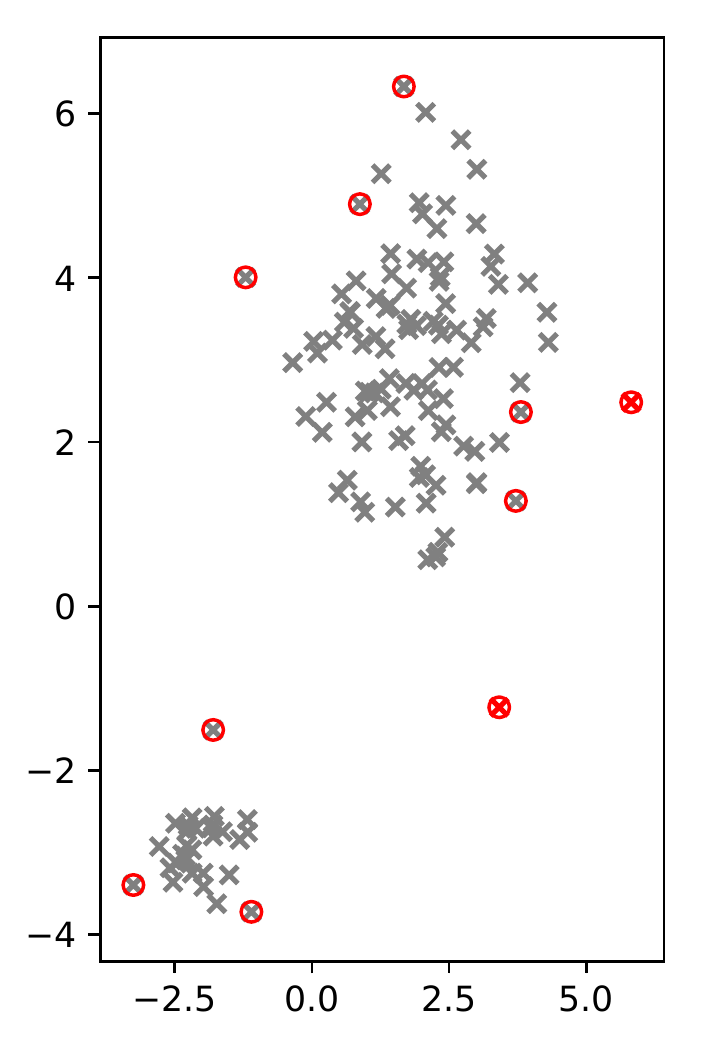}}\quad
  \subfigure[]{\includegraphics [width=0.25\textwidth]{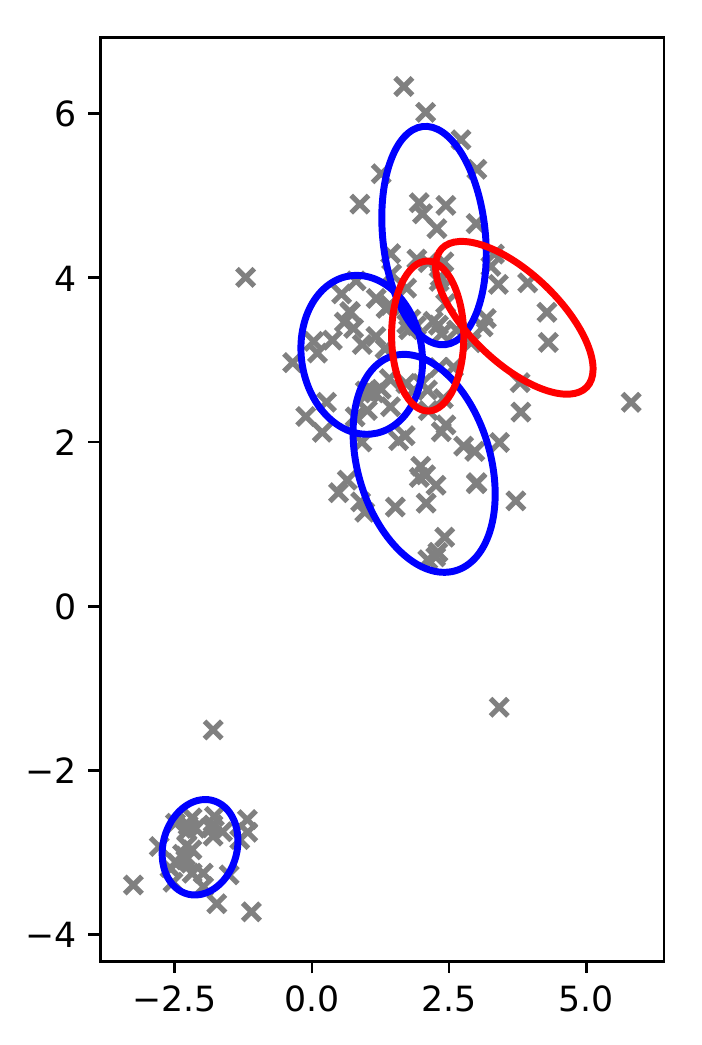}}\quad
  \subfigure[]{\includegraphics[width=0.27\textwidth]{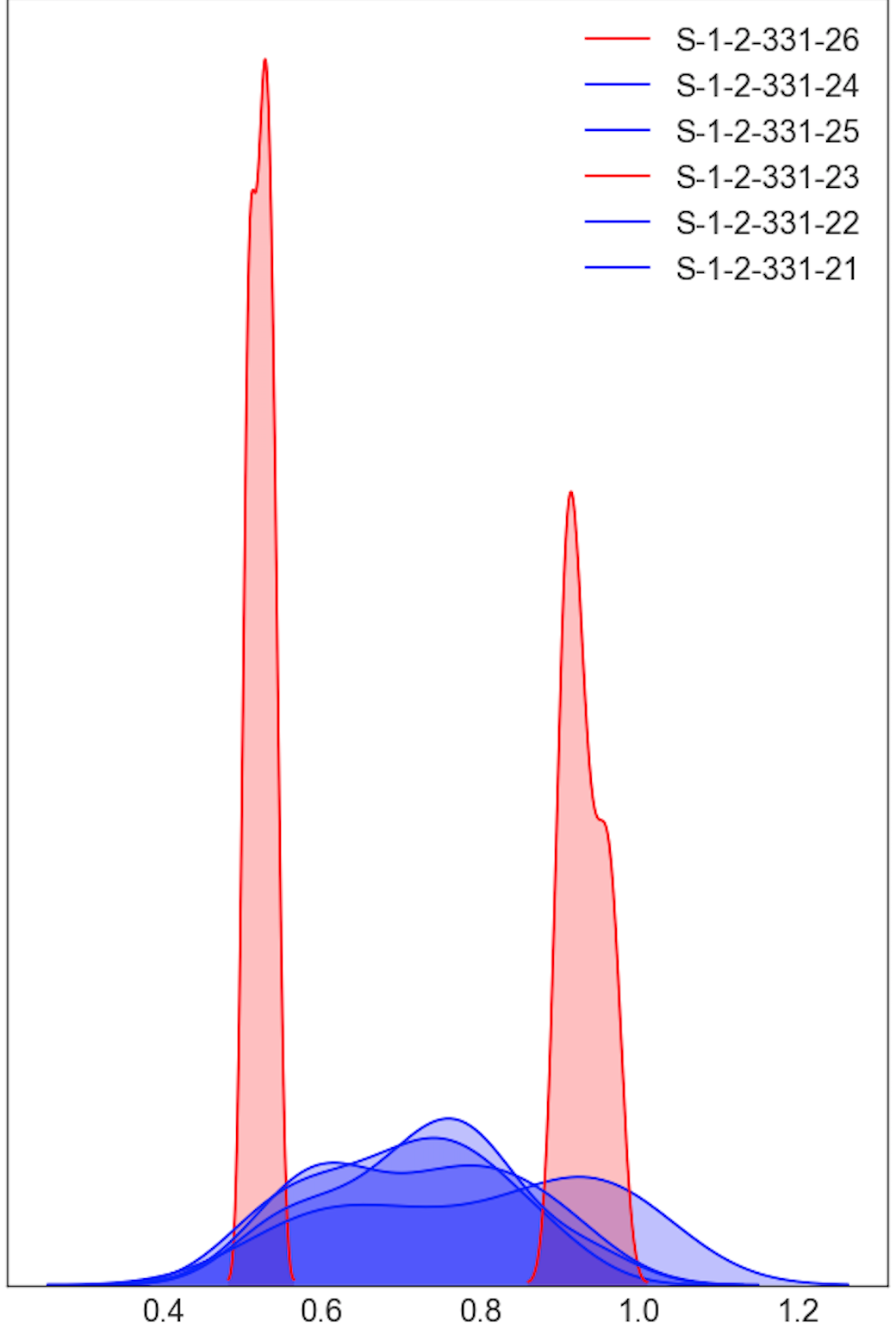}}
  \caption{
  (a) Feature space of  6 sessions projected in 2-D latent space, point anomalies are marked in \textcolor{red}{circles}
  (b) Known Group membership of features per session  \textcolor{red}{Attacker} session, \textcolor{blue}{normal/user} session.
  (c) Latent  distribution $Z$ of 6 sessions, \textcolor{red}{Attacker} sessions and  \textcolor{blue}{normal/user} session.}
  \label{concept}
\end{figure*}

On the defensive side, the detection of patterns that differ from typical behavior is utterly important to detect new threats. 
This requirement has been satisfied by using algorithms that are capable of detecting point anomalies. 
Many of such approaches cannot detect 
a variety of different deviations that are evident in group datasets. 
For example, the activity of a domain admin on a machine can be similar to an attacker activity confusing any point anomaly detectors. Identifying attacker activities, in this case, require more specialized techniques for robustly differentiating such behavior. 

Group Anomaly Detection (GAD) aims to identify groups that deviate from the regular group pattern \cite{gad_survey}. There are many intresting applications of GAD in various domains: (a) discovery of Higgs bosons in physics \cite{OCSMM};  anomalous galaxy clusters in
astronomy \cite{OCSMM}; unusual vorticity in fluid dynamics\cite{GLAD}; prevent disease outbreaks and  \cite{Weng:2002}; 

To the best of our knowledge, GAD has not been applied to solve security problems we have addressed in this work.

In this paper, a Windows session -- User's Security Identifier (SID) -- is used as the grouping key and all activities occurring in that session are attributed to group behavior.
Figure \ref{concept} illustrates an example of how group anomaly detection can be applied to discover attacker activity. 
For example, out of six sessions observed on an endpoint (see Table \ref{tab:command_session}, four are normal behavior and two sessions are of real attacker activity. 
Existing state-of-the-art point anomaly detectors (Figure \ref{concept}-(a)), miss the attacker activity as the attacker's session does not exhibit any outlier characteristics when compared to normal activities. 
When we apply distribution (Figure \ref{concept}-(c)) based group anomaly method, with carefully chosen grouping (Figure \ref{concept}-(b)) and deviance functions (Equation \ref{eq:groupreference}), it clearly differentiates between benign and attacker sessions.

The main contributions of this paper are as follows: 
\begin{itemize}

\item We apply GAD to identify anomalies and propose a new deep generative model: Adversarial Autoencoders-$\alpha$ (AAE-$\alpha$) based on Adversarial Autoencoders (AAE) to detect attacker activities.
Although anomaly detection has been applied to solve multiple problems in the security domain, discovering advanced attackers leveraging off-the-shelf tools and system applications in their campaigns has not been thoroughly explored for real world datasets.

\item We build two novel deep learning embedding models: \textit{Behavior2Vec} and \textit{Command2Vec} for feature extraction specific to the security domain. 
The low-level behavior observed on an endpoint is mapped to a subset of MITRE ATT\&CK tactics that are used as features to build the models. 

\item The method was tested on real world data collected from 12000 enterprises with more than 100k endpoints for a period of two months. 
The empirical results demonstrate that our approach is effective and robust in detecting attacks including targeted attacks, pen-tests, and attack campaigns leveraging custom tools. 

\end{itemize}

The rest of the paper is organized as follows. 
We review related work in Section~\ref{sec:related}.
We describe the methodology and elaborate on our proposed solution in Section~\ref{sec:method}. 
Description of datasets, feature extraction methods, and motivations are presented in Section~ \ref{sec:dataset}. 
In Section ~\ref{sec:results}, we present the results of our experiments.
Finally, Section~\ref{sec:conclusion} concludes our work and gives some recommendations for further research.

\section{Related Work}
\label{sec:related}

The field of anomaly detection has been thoroughly studied since decades \cite{surveyrecent,Akoglu2015,hodge2004survey, chandola2009anomaly}. 
The usefulness of such algorithms to host protection has been confirmed long-time ago \cite{Denning1987} as this is the main tool to detect previously unknown attacks. 

Several approaches use log correlation to perform detection by clustering similar logs and by identifying causal relationships between logs \cite{steven2004,wei}. 
Beehive \cite{beehive} identifies potential security threats from logs by unsupervised clustering of data-specific features, and then manually labeling outliers. 
Oprea \cite{opera} uses belief propagation to detect early-stage enterprise infection from DNS logs. 
BotHunter \cite{bothunt} employs an anomaly-based approach to correlate dialog between internal and external hosts in a network. 
HERCULE \cite{Kexin} leverages community discovery techniques to correlate attack steps that may be dispersed across multiple logs. 
All the aforementioned techniques either find point anomalies or group similar logs by clustering. 
In our approach, we discover attacker activity that is similar to normal activity and not a point anomaly. 

Recent advances in Recurrent Neural Networks, motivated additional work in this field, for example, Du et al.~\cite{Du17} proposed DeepLog that uses Long Short-Term Memory (LSTM) networks. 
This approach is able to capture high dimension non-linear dependencies. 
In contrast to our work, their approach uses raw system logs that contain unstructured information, typically in an unordered manner, due to concurrency reasons; and also, the idea has been used to detect only one type of threat: Denial of Service (DoS). 
The problem has been addressed by Brown et al.~\cite{Brown18} who performed unsupervised anomaly detection by extending a typical LSTM architecture with attention mechanism. 
The approach provides an output if the input log is malicious or not, without specifying generated threat category.

A recent survey by Toth et al. \cite{gad_survey} on group anomaly detection and group change detection describes developments in the application of GAD in static and dynamic situations. 
Xiong \cite{Collective} provides a more detailed description of current state-of-the-art GAD methods. 
Yu et al.~\cite{SurveySocialMedia} further review GAD techniques, where group structures are not previously known. 
However, clusters are inferred based on additional information of pairwise relationships between data instances. Recently, \cite{imagesgroup} Chalapathy  et al. proposed autoencoder based  group anomaly detection technique for images.

\section{Methodology}
\label{sec:method}

A group is a collection of two or more related data instances \cite{gad_survey}. Groups which deviate significantly when compared with other groups are known as group anomalies. Group anomalies can be point based or distribution based. Point-based anomalous groups are where all members are also point-wise anomalies. In a distribution-based group anomaly, a collection of points differs from expected group patterns; however, individual data instances may not seem anomalous. 
We formulate the problem of discovering attacker activity that may be  similar to normal admin/user activity as a distribution based group anomaly detection. Attacker behavior may be similar to normal admin user, but when we compare across groups of admins/users activities they significantly deviate by their distributions. 
To capture group behaviors, we need an objective function that minimizes intra-group variations while simultaneously achieving maximal inter-group separation.

First, consider a set of groups $\mathcal{G} = \big\{  {\bf G}_m \big\} _{ m=1 }^M  $, where the $m$th group contains $N_m$ observations with  

\begin{equation}
{\bf G}_m = \big(X_{nv}\big) \in \mathbb{R}^{N_m \times V} 
\end{equation}

where $X_{nv}$ is the vth feature (v = 1, 2, ..., V) of observation n (n = 1,2, .., $N_{m}$) in the group $G_{m}$, R is a continuous value domain. The total number of individual observations is $N=\sum_{m=1}^M N_m$.

For inter-group similarity, we measure the distance between groups by  restricting the group boundaries by $\alpha$, The parameter $\alpha$ in Eq.~\ref{eq:groupreference} determines the group spread and helps to achieve uniform inter-group variations. Our visualisation of the learnt features in Fig.~\ref{concept} demonstrate that the attacker activity mainly falls in the tail end of distributions. For our proposed method, normal activity in groups is directly related with the value of parameter $\alpha$. 
We perform experiment on  dataset for different values of the the parameter $\alpha=\{1,20\}$. 
\begin{align}\label{eq:groupreference}
\mathcal{G}^{(exp)}  = max \Big[
\exp\Big(-\frac{\parallel\textbf{\bf G}_i-\textbf{\bf G}_j\parallel^2}{\alpha}\Big)\Big] 
\end{align} 
The results in Fig.~\ref{fig:auc_aplha} show that the optimal performance is achieved for values of $\alpha$  between $10$.

To learn intra-group compactness of features we design a loss function
\begin{align}\label{eq:grouploss}
    L_{gg} = \sum_{z} \max \big(0, \lambda + d(\textbf{f}_i, \textbf{G}_z) - d(\textbf{f}_i, \textbf{G}_{y})\big) \; : z \neq y,
\end{align}

where $\textbf{f}_i$ is activity feature observed  in a group $\textbf{G}_z$,
$d(\textbf{f}_i, \textbf{G}_z)$ is the similarity of the  $\textbf{f}_i$  with other activity features in the same group, $d(\textbf{f}_i, \textbf{G}_{y})$ is  similarity of the activity feature  $\textbf{f}_i$  with other groups represented by $\textbf{G}_{y})$, and $\lambda$ is the enforced margin.

Then, a distance metric $d(\cdot,\cdot) \ge 0 $ is applied to measure the deviation of a particular group from the other groups. 
The distance score $d\Big(\mathcal{G}^{(exp)} , {\bf G}_{m} \Big )$ quantifies the deviance of the $m$th group from the expected group pattern, where larger values are associated with more anomalous groups. 
In summary the proposed method provides us: \textbf{(a)} the  adjustability  to enforce  margin   constraints on group distributions, \textbf{(b)} capture  exact group boundaries for attacker and normal activity, and \textbf{(c)} control the variance of learned features in a group  and therefore enhancing intra-group compactness.   

\begin{algorithm}[htp]
\caption{Group anomaly detection using AAE-$\alpha$}
\label{algo:gadVae}

\DontPrintSemicolon
\SetAlgoLined
\SetKwInOut{Input}{Input}\SetKwInOut{Output}{Output}
\Input{ Groups $\mathcal{G} = \big\{  {\bf G}_m \big \} _{ m=1 }^M $}

\BlankLine
\Output{Group anomaly scores \textbf{S} for input groups $\mathcal{G}$ }
\BlankLine
  \Begin
  {
      
        {
            
            {
          
          Draw a random latent representation $z-z\_tail \sim f_\phi(z|\mathcal{G})$                         }
         Reconstruct sample using decoder $   g_\psi(\mathcal{G}|z)$\;   
          \For{(m = 1 to M)}{
               Compute the score
        ${ s}_m =d\Big(\mathcal{G}^{(exp)}, {\bf G}_{m}   \Big ) $

        \BlankLine
    }
     Sort scores in descending order  
     \textbf{S}=$ \{s_{(M)} >\dots>s_{(1)} \}$\;
     $\{s_{(m)} \}_{m=1}^M  $
    Groups that are furthest from $\mathcal{G}^{(ref)}$ are more anomalous.\;   
        }
        \textbf{return S}
        
}

\end{algorithm}

\noindent \textbf{Adversarial Autoencoders (AAEs).}
An AAE is a generative model that is trained with dual objectives (i.e., a traditional reconstruction error criterion and an adversarial training criterion~\cite{makhzani2015adversarial}). The encoder in AAE learns to convert the data distribution to a latent representation with an arbitrary prior distribution, attempting to minimize the reconstruction error. In other words, a GAN is attached to the latent layer. 

We specifically choose an AAE over other variants in our work for the following reasons:
\begin{itemize}
    \item  In order to capture the true distribution of the latent space, we want to filter out latent vectors near the tail-end of the latent distribution in the scoring logic. 
    For other GAN variants, it is non-trivial to encode an arbitrary sample back into the latent space. 
    \item Since session data follows seasonal patterns with AAEs, we can capture a variety of prior distributions on the latent space without knowing the exact functional form in advance. 
\end{itemize}
We calculate the magnitude of each encoded latent vector as measured from the latent mean and filter vectors with magnitudes below the $\alpha$ percentile for calculating inter group similarity as described in Equation \ref{eq:groupreference}.
In our system, we set  $\alpha$ to 10th percentile norm of the training set vectors, as it showed the most robust behavior throughout all experiments. One can also fix the $\alpha$ , e.g., to a specified number of standard deviations, without optimizing on the training set.

The process of how we calculate group anomaly score is further explained in Algorithm \ref{algo:gadVae}. Group anomalies are effectively detected when functions $f$ and  $g$ respectively capture properties of group distributions and appropriately combine information into a group reference.


\noindent{\textbf{Training.}} Activities are recorded by the agent installed on the endpoint. These activities include process, file and command creation, termination and executions. 
We use concatenation of feature vectors of the group as the input layer in the AAE  specifically, $\bm{X}_m = \bf {concat}(\bm{x}_{1}, \dots, \bm{x}_{n})$ where $\bm{x}_{i}$ is ${i}$th feature vector of a group.

The AAE-$\alpha$ architecture is trained according to the loss function given in Equation  (\ref{eq:grouploss}). 
 Given known group memberships, AAE is fully trained on input groups to obtain a representative group expectation score $\mathcal{G}^{(exp)}$. Grid Search was used for hyper-parameter tuning,  including the number of hidden-layer nodes $H \in \{120, 60, 320\}$, and regularization  $\lambda$ within range ${[0, 100]}$. 
The output scores are sorted according to descending order, where groups that are furthest from  $\mathcal{G}^{(exp)}$  are considered anomalous.

\section{Feature Extraction and Pre-processing}
\label{sec:dataset}

In our system, a group is defined by a set of activities observed in a single windows session. 
Session activity consists of signature names, command line text, session properties, file, and  process information. We derive multiple features including density-based feature, session property based features, static, dynamic, reputation and prevalence based features from file and process information. 
For a given day, We group all raw events by session ID field in the dataset. Next, we extract relevant features from each group.
Below we give a brief overview of our feature extraction pipeline.

\textbf{Command2Vec.} 
An Agent installed on the endpoint matches a set of signatures/rules based on the activity and generates a log. 
Logs contain the signature name, timestamp, command line text executed by a different process in a given session. 
At training time, we group all command lines by signature name from the data set. 
Next, we use command line text to train a sentence2vec model\cite{sent2v} with 200-dimensional sentence embeddings, with window size 20 and by setting negative examples to 10 for each signature. 
Finally, at feature extraction time for a given session we (a) concatenate all command line text for each signature; (b) use the signature model to extract vectors for the concatenated text; (c) adopt a 3 layer autoencoder on the all the vectors to
reduce the dimension.
For example, let us assume in a given session, $Session1$ can be denoted as a group of signatures and command lines. 
$$Session1[
[s_{1},[c_{s1_1}..c_{s1_n}],
[s_{2},[c_{s2_1}..c_{s2_n}]
..[s_{m},[c_{sm_1}..c_{sm_n}] ]$$ 

We concatenate command line text $[c_{s1_1}..c_{s1_n}]$ as one blob of text and feed it to already trained signature model to extract feature vectors. 
Each signature model generates 200 x 1 vector if in a session we have M signatures then the final vector length will be 200 x M. 
Now, this 200 x M is fed into a 3 layer autoencoder to reduce the dimensionality from  200 x M to 200 x 1.

\begin{figure}[htp]

     \includegraphics[width=\columnwidth]{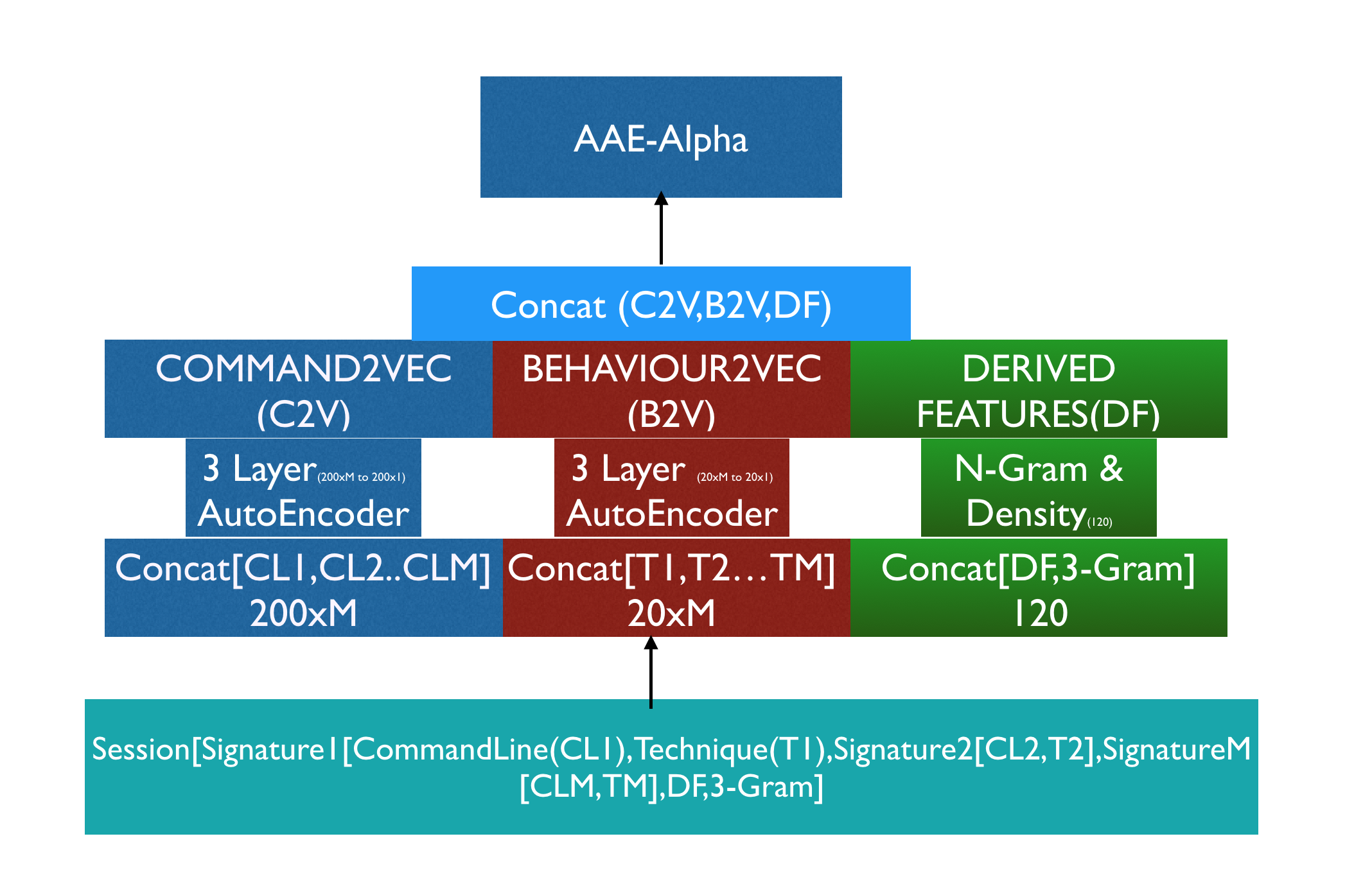}
      \caption{Feature extraction pipeline: (a) For a given session- aggregate all signatures and commandline text and extract embeddings; (b) Run dimensionality reduction on aggregated embeddings to reduce to size 200x1 and 20x1; (c) Feed the resulting embeddings to AAE network as features.}
       \label{fig:features}
\end{figure}

\textbf{Behavior2Vec.} 

MITRE's ATT\&CK framework\cite{mitre} is a community project to capture adversary  Tactics, Techniques, and Procedures (TTP) from real world intrusions. Each TTP or a combination of TTP helps to achieve high-level goal of an attacker. 
 
First, we map the AV signature rules to high-level behaviors manually. Table \ref{tab:mitre} provides an overview of mappings between signature rules to high-level techniques from  MITRE's ATT\&CK framework \cite{mitre}. 
We group all logs for a given session ordered by timestamp and collect all the signatures in the session. 
Next, we create a sequence of MITRE techniques for a given session by looking up from \ref{tab:mitre}. 
We use these sequences to train a sentence2vec model with 20-dimensional sentence embeddings with window size 2 and by setting negative examples to 2. 
The advantage of high-level mapping is two-fold: (a) we can communicate to the human analyst in an intuitive way instead of just black box anomaly score; (b) we use these mappings to build a sentence2vec model for feature extraction.

For strings such as (a) File, process and directory names and paths; (b) File Signer Information(Subject, Issuer); (c) Process thread names, function names we use tri-gram model for feature extraction. 
Figure \ref{fig:features} summarizes feature extraction pipeline.  

\begin{table}[!htp]
 \caption{High-level behaviors mapped to a subset of MITRE techniques.}
 \label{tab:mitre}
\begin{tabularx}{\linewidth}{ |X |X|}
\hline
\textbf{High-level Behaviors} & \textbf{Technique}\\
\hline
Bits Admin activity  & BITSJobs \\ 
\hline 
Command launches by \textit{msiexe, WMI, powershell}, network activity by command line  & Command-LineInterface \\
\hline 
Memory access of \textit{LSASS, CryptDll}   & CredentialDumping and LSASSDriver \\
\hline 
Query for Security tools, Registry changes to trusted process  & DisablingSecurityTools \\
\hline 
Password dumping and Key logging activity & ExploitationforCredentialAccess \\ 
\hline 
Random folder creation, Proxy changes, Registry and startup folder changes  & HiddenFilesandDirectories \\
\hline 
WMIC activity  & WMIC  \\ 
\hline 
 WScript runs, Powershell launches other process  & Scripting \\
\hline 
Build tools, script execution by common tools with network traffic  & TrustedDeveloperUtilities \\
\hline 
Scheduled tasks added or launched   & ScheduledTask \\
\hline 

Process injections with network activity & ProcessInjection \\
\hline 
Client tools  with network activity other then browsers  & ExploitationforClientExecution \\
\hline 

Suspicious Browser Helper Object modification KeyLogger activity,
Suspected screen capture attempted & Collection 
\\ \hline 
\end{tabularx}
 
 \end{table}
 
\textbf{Density based Features.} 
These features capture the inherent properties of a session (group), for example, how popular a session is in the enterprise, type of activity in that group - backdoor, attacker, helpdesk, remote admin. 
For a given session, we extract (a)number of machines on which the session was seen; (b)behaviors which are   shared across other sessions; (c)command lines shared across other sessions.
A system admin may use same commands across multiple endpoints to update software or apply a patch while using the same user account. Some attackers may also reuse the same session across multiple organizations to conduct their campaigns.

\textbf{Session Features.} 
Typically, targeted attackers reuse, migrate or create a new user or session for lateral movement and data exfiltration. 
Session properties like (a) remote vs local; (b) admin vs non-admin; (c) user was idle vs active give insights into the type of user activity in a given session.

\textbf{Static and Dynamic Features.}
Static and dynamic features of the files and process executing in a given session help us understand the type of file or process used to create a registry entry and from which place in the file system. 
Typically, attackers use temporary folders in the file system for malicious file downloads and use schedule tasks and registry entries for persistence. We extract (a)Signer information of the file; (b)  
Entropy; (c)Import functions and Sections; (d)File Header information; (e)Export functions used; (f)File  Paths, and directory; (g) Process names.

\textbf{Reputation and Prevalence based Features.}
Popularity and reputation of file or process inside an enterprise and across enterprises help us capturing the toolset of the attackers. Attackers reuse the tools in multiple attack campaigns. 
Reputation and prevalence of a file change over time, so in our system, we use histograms collected over a period of 60 days as one of the features. Histogram features include (a)Prevalence of file, and process across population and inside an enterprise; (b)File signer subject reputation; (c)File reputation. We want to note that this data was provided by Anti-Virus firm for all the file hashes  in the dataset.    

\section{Experiments and Results}

\label{sec:results}
In this section, we discuss the results of experiments for the proposed method.

\subsection{Datasets}

The dataset consists of behavioral events (e.g., registry value changes, process executed, command lines responsible for the activity, and windows session IDs occurring on endpoints of enterprise customers for two month time period. 
A large antivirus company that opted in to share their data, such that through large-scale data analysis new methodologies can be developed to increase the existing advanced threat detection capabilities. 
To protect customer identities, all sensitive information is anonymized.
The behavioral events are collected from more than 12000 enterprises with a total of 100k endpoints, which is only a subset of data processed by the antivirus company. 
Even when this data set is small our results show that it is sufficient to model attacker activity. 

The fields we are particularly interested in are: (a) (anonymized) machine and customer identifiers; (b) static features of file or process; (c) reputation and prevalence of the file or process in enterprise; (e) command lines executed by process/file; (f) file name and directory; (g) timestamps (in UTC) of  activity; (h) low level behaviors. 
Table \ref{tab:dataset} gives an overview of different statistics of the dataset. 

\begin{table}[htp]
\caption{Statistics of the datasets.}
    \label{tab:dataset}
    \centering
    \begin{tabular}{|c|c|} \hline
        \textbf{Type}  & \textbf{Unique counts}\\ \hline
 Sessions  & 5998744\\ \hline
 Known Malicious Sessions   & 926 (0.0154\%)\\ \hline
 Command lines  & 3004454\\ \hline
 Signatures & 992\\ \hline
 Endpoints & 100233 \\ \hline
  Enterprises & 12000 \\ \hline
  Features Extracted per session & 340 \\ \hline
    \end{tabular}
\end{table}

We compare our approach with both point-wise anomaly detectors and group anomaly detectors, which we briefly describe below.

\begin{description}[style=unboxed,leftmargin=0cm, parsep=10pt]

\item[One Class Support Vector Machines (OC-SVM).] \cite{ocsvm1999} are kernel based boundary based anomaly detection techniques  which identify decision boundary around normal examples. 
The $\nu$ parameter is set to the expected anomaly proportion in the dataset, i.e., 0.00015, which is assumed to be known, whereas the $\gamma$ parameter is set to $1/m$ where $m$ (340) is the number of input features. 

\item[Isolation Forests (IF).] \cite{iforest2008} is a partition-based method it first builds tress to randomly split values across chosen features. Scikit-learn \cite{scikit-learn} package default parameters were used   in our experiments. 
 
\item[Deep Autoencoding Gaussian Mixture Model (DAGMM).] \cite{zong2018deep} is a state-of-the-art autoencoder based method for anomaly detection. It trains a auto encoder to use its latent representations, reconstruction error to feed into estimation network to a  Gaussian Mixture Model (GMM) to predict anomalous data point.
For the experiments, compression network with 3 dimensional input to the estimation network, where one is the reduced dimension and the other two are from the reconstruction error, the compression network runs with FC(340,120), FC(120, 60, tanh)-FC(60, 30, tanh)-FC(30, 10, tanh)-FC(10, 1, none)-FC(1, 10, tanh)-FC(10, 30, tanh)-FC(30, 60, tanh)-FC(60, 120, tanh)--FC(120, 340, tanh), and the estimation network performs with FC(3, 10, tanh)-Drop(0.5)-FC(10, 4, softmax).

\item[Mixture of Gaussian Mixture Models (MGMM).] \cite{MGM} assumes that data generated data follows Gaussian mixture
where anomalous group does not follow regular mixture of features.
The number of regular group behaviours $T$=1 and number of Gaussian mixtures $L$=4 was used for training.

\item[One Class Support Measure Machines (OCSMM).] \cite{OCSMM} 
uses  discriminating method to discover  between  regular  and  anomalous  group behaviours using a  hyperplane.  OCSMM maximises the margin between two classes  separated  based  on  the  hyperplane and learns  patterns  from  one-class  that  exhibits  the  dominant  behavior  in a  dataset.

\item[Deep generative models (VAE and AAE).] VAE\cite{Kingma2013} and AAE \cite{makhzani2015adversarial} are generative variants of auto encoder which use reconstruction probabilities~\cite{an2015variational} to compute anomaly scores.
AAE relaxes this constraint by using a GAN as described in Section \ref{sec:method}. 
Both AAE and VAE used four layers of (conv-batch-normalization-elu) in the encoder part and four layers of (conv-batch-normalization-elu) in the decoder network. 
The AAE network parameters such as (number of filters, filter size, strides) are chosen to be (340,3,1) for first and second layers and (120,3,1) for third and fourth layers of both encoder and decoder layers. 
The middle hidden layer size is set to be same as rank $K = 60$ and the model is trained using Adam~\cite{kingma2014adam}.
The decoding layer uses sigmoid function in order to capture the nonlinearity characteristics from latent representations produced by the hidden layer. 

The proposed method AAE-$\alpha$ uses similar architecture as standard AAE but instead of drawing samples from the real distribution of data we filter 10th percentile to remove outliers. 
Figure \ref{fig:auc_aplha} shows the sensitivity of AUC with percentile value. In our experiments, we used $\alpha$=10.  

\begin{figure} [htp]
    \centering
    \includegraphics[width=.8\columnwidth]{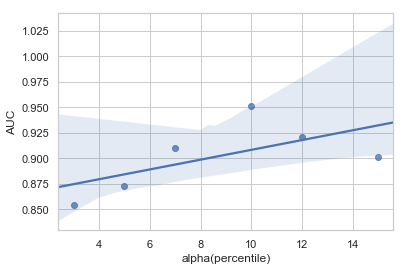}
    \caption{Model AUC sensitivity with $\alpha$.}
    \label{fig:auc_aplha}
\end{figure}

\end{description}

The dataset provided to us was not labeled, but a small set of sessions which had  one attacker group activity identified  by  analyst in the past were shared with us. We used these sessions as validation dataset, machine identifiers found in these sessions were also filtered from the dataset to make sure we evaluate the models on completely unseen data.

The performance of the model is evaluated using the area under the precision-recall curve (AUPRC) and area under the ROC curve (AUROC) metrics ~\cite{Davis:2006}.
Table \ref{tbl:syn} summarizes the AUROC and AUPRC values of different methods. 
Interestingly, we observe that AAE and VAE  methods achieve similar results when compared with each other, sampling data filtered by $\alpha$ percentile increases performance for the given dataset.

\begin{table}[!htp]
\caption{Results for  different methods on the dataset. 
The highest performances are in gray.}
\centering
    \label{tbl:syn}
   
    \begin{tabular}{|c|c|c|}
       
        \hline
       
        \textbf{Method} &\textbf{AUPRC} & \textbf{AUROC}
          \\
        \hline
     OCSVM&0.6212 &0.4508 \\ \hline
     DAGMM&0.8422 &0.5100 \\ \hline
     IF&0.6977 &0.4833 \\ \hline
     AAE & 0.9080  & 0.5130 \\ \hline
     VAE & 0.9001 & 0.5007  \\\hline
     MGM& 0.8584 & 0.4809 \\\hline
     OCSMM&0.8233 &0.5097\\\hline
     AAE-$\alpha$ (Ours) & \cellcolor{gray!25}{0.9526} & \cellcolor{gray!25}{0.6022}\\\hline
     
    \end{tabular}

\end{table}

\subsection{Discussion}
\begin{description}[style=unboxed,leftmargin=0cm, parsep=10pt]

\item[New attacks discovered.]
We filtered all known malicious sessions and corresponding machine identifiers from the dataset.
Next, We run the model to assign anomaly scores to all sessions in the dataset and sort the sessions by scores and choose top 0.001\% (around 6000 sessions which were seen in 4000 enterprises). 
We presented top 0.001\% anomalous sessions scored by model to threat analysts for review. They observed (a) 1260 sessions with red teaming and, pen testing activity which have similar behavior of an attacker; (b) 330 sessions have the activity of malicious banking trojan which steals sensitive information using  off the shelf tools; (c) 40 sessions with 2 new targeted attack groups tool activity; (d) 2833 sessions with malicious executions of tools like PowerShell, WMIC, FTP, and client tools; (e) 620 sessions with security  tool scans for checking malicious activity; (f) 417 sessions with benign activity like logging remotely and installing updates.

\item[False Positives.] 
We noticed 417 false alarms detecting security operation tasks and remote sessions as malicious. One of the reasons for higher scores for these sessions is because these sessions were shared across multiple organizations. 
Since all the data is anonymized, we speculate that this may be the case of a contractor or part-time employee working for multiple organizations or it may be a feature of a  security tool. 
Another set of false positives comes from scanning activity of security tools. 
These may be reduced by whitelisting rules.

\item[Anomaly Scores interpretation and Alerting.]  
Although research on anomaly detection for cyber defense spans more than two decades, adoption of statistical methods is limited due to two main reasons: (a) high false positive rate; (b) uninterpretable alerts. 
Analysts are inundated with a large number of alerts and triaging them takes significant time and resources; this results in low tolerance for false alarms and alerts that provide no contextual information to guide the investigation. 
To address this issue, we map raw event data into a high-level abstraction of  MITRE ATT\&CK TTPs. 
This helps the analyst to interpret the score with the technique name. 
We also key all raw data with corresponding session ID, machine identifier, and timestamp. Sessions with high anomaly scores can be presented to an analyst for further review.

\end{description}
\section{Conclusions and Future work }
\label{sec:conclusion}
We developed a group anomaly detection model using adversarial autoencoders to detect targeted attackers who hide their activity using dual-use tools and system installed applications. 
We build two deep learning models one for feature extraction and one for generating anomaly score. 
We map low-level behaviors to high-level MITRE ATT\&CK classification, which helps human analysts to interpret anomaly scores.

We tested our method on the real world dataset collected for two month time period from 12k enterprises. 
We discovered around 5000 instances of malicious activity -- 40 sessions with custom tool activity of 2 targeted attack groups, around 3000 sessions with malicious activity using system installed tools, and 1000 sessions with pen testing and red teaming activity without any labeled data and supervision. 
The results show that our method successfully detects threats that hide in plain sight with high precision and low false alarm rates.

Despite some interesting results of our proposed method to discover advanced threats, we want to highlight its inherent limitations and subjects for future work. 
The visibility of activities occurring on the endpoints is limited by collector configuration and settings. We may have missed some activity due to agent throttling settings and reporting configurations.
We plan to address this limitation by augmenting the dataset with other data sources like email data and network data. 
As part of future work, we also aim to test the model performance on a long time ranges, create a multi-class classifier to identify multiple attack groups and evaluate on audit logs from other operating systems like Mac and Linux.

\end{document}